\DocumentMetadata{}
\documentclass[acmsmall,screen]{acmart}

\usepackage{booktabs}
\usepackage{multirow}
\usepackage{graphicx}

\usepackage{amssymb}
\usepackage{amsmath}
\usepackage{amsfonts}

\usepackage[most]{tcolorbox}

\newtcolorbox{findingbox}{
  enhanced,
  colback=gray!15,
  frame hidden,
  boxrule=0pt,
  sharp corners,
  left=8pt,
  right=8pt,
  top=4pt,
  bottom=4pt,
  borderline west={3.2pt}{0pt}{black}
}

\begin{document}

\title{Can Vision-Language Models Handle Long-Context Code? An Empirical Study on Visual Compression}



\author{Jianping Zhong}
\affiliation{%
  \institution{Zhejiang University}
  \country{Ningbo, China}}
\email{jpz@zju.edu.cn}

\author{Guochang Li}
\affiliation{%
  \institution{Zhejiang University}
  \country{Hangzhou, China}}
\email{gcli@zju.edu.cn}

\author{Chen Zhi}
\authornote{Chen Zhi is the corresponding author.}
\affiliation{%
  \institution{Zhejiang University}
  \country{Ningbo, China}}
\email{zjuzhichen@zju.edu.cn}

\author{Junxiao Han}
\affiliation{%
  \institution{Hangzhou City University}
  \country{Hangzhou, China}}
\email{hanjx@hzcu.edu.cn}

\author{Zhen Qin}
\affiliation{%
  \institution{Zhejiang University}
  \country{Ningbo, China}}
\email{zhenqin@zju.edu.cn}

\author{Xinkui Zhao}
\affiliation{%
  \institution{Zhejiang University}
  \country{Ningbo, China}}
\email{zhaoxinkui@zju.edu.cn}

\author{Nan Wang}
\affiliation{%
  \institution{Shenzhou Aerospace Software Technology Company Limited}
  \country{Beijing, China}}
\email{wangnan02026@163.com}

\author{Shuiguang Deng}
\affiliation{%
  \institution{Zhejiang University}
  \country{Hangzhoum China}}
\email{dengsg@zju.edu.cn}

\author{Jianwei Yin}
\affiliation{%
  \institution{Zhejiang University}
  \country{Hangzhou, China}}
\email{zjuyjw@cs.zju.edu.cn}



\begin{abstract}
Large Language Models (LLMs) struggle with long-context code due to window limitations. Existing textual code compression methods mitigate this via selective filtering but often disrupt dependency closure, causing semantic fragmentation. To address this, we introduce LongCodeOCR, a visual compression framework that renders code into compressed two-dimensional image sequences for Vision-Language Models (VLMs). By preserving a global view, this approach avoids the dependency breakage inherent in filtering. We systematically evaluate LongCodeOCR against the state-of-the-art LongCodeZip across four benchmarks spanning code summarization, code question answering, and code completion.

Our results demonstrate that visual code compression serves as a viable alternative for tasks requiring global understanding. At comparable compression ratios ($\sim$1.7$\times$), LongCodeOCR improves CompScore on Long Module Summarization by 36.85 points over LongCodeZip. At a 1M-token context length with Glyph (a specialized 9B VLM), LongCodeOCR maintains higher accuracy than LongCodeZip while operating at about 4$\times$ higher compression. Moreover, compared with LongCodeZip, LongCodeOCR drastically reduces compression-stage overhead (reducing latency from $\sim$4.3 hours to $\sim$1 minute at 1M tokens). Finally, our results characterize a fundamental coverage--fidelity trade-off: visual code compression retains broader context coverage to support global dependencies, yet faces fidelity bottlenecks on exactness-critical tasks; by contrast, textual code compression preserves symbol-level precision
while sacrificing structural coverage.
\end{abstract}
\maketitle

\section{Introduction}
In recent years, Large Language Models (LLMs) for code have made substantial progress on tasks such as code completion\cite{guo2023longcoder,liu2023repobench}, program synthesis\cite{zhang2023algo}, program repair\cite{li2025swe,chen2025swe}, unit test generation\cite{chen2024chatunitest} and LLM-based application analysis\cite{wu2025llmapphub}, significantly advancing the development of intelligent developer assistant tools.
As application scenarios evolve from single-file scopes to repository-scale, models increasingly need to resolve cross-file evidence, including API definitions, complex call relationships, and configuration scripts.
Such critical signals are often scattered across multiple files and locations, causing the required input context to frequently grow to tens of thousands of tokens or more. However, feeding such long contexts directly into LLMs creates a fundamental tension between information integrity and computational feasibility. Specifically: (1) under the standard Transformer architecture, the computational complexity of the self-attention mechanism\cite{vaswani2017attention} grows quadratically with sequence length, leading to prohibitive memory footprints and inference latency; (2) in long and noisy inputs, evidence localization becomes unreliable\cite{li2024loogle}, a phenomenon known as ``lost-in-the-middle''\cite{liu2024lost}; and (3) contexts may still exceed hard window limits, where forced truncation disrupts the dependency closure of the code, thereby leading to reasoning failure.\cite{bogomolov2024long}

However, there has been limited progress in addressing long context limitations for code. General text compression methods like LLMLingua\cite{jiang2023llmlingua} and Selective Context\cite{li2023compressing} fail to account for code-specific characteristics and often break syntactic validity. Retrieval-augmented generation (RAG)\cite{cheng2024xrag} may overlook implicit dependencies within the context. Existing code compression methods, such as DietCode~\cite{zhang2022diet}, SlimCode~\cite{wang2024natural}, and LongCodeZip~\cite{shi2025longcodezip}, compress long contexts by selectively pruning or filtering less useful code snippets. By relying on binary keep-or-discard decisions, these methods inevitably fragment the context. More fundamentally, these existing methods can be unified under a single paradigm—textual code compression. This classification arises because they primarily operate on one-dimensional token sequences via selective filtering.

Meanwhile, vision-text compression (VTC), also known as contextual optical compression, has emerged as a promising cross-modal paradigm for mitigating context limitations recently.
By rendering long documents into compressed two-dimensional image sequences, VTC replaces token-level pruning with visual representations under a fixed visual token budget, as explored in DeepSeek-OCR~\cite{wei2025deepseek}, AgentOCR\cite{feng2026agentocr} and Glyph~\cite{cheng2025glyph}.
This approach avoids the dependency breakage inherent in selective filtering by preserving a global view of the code context.
However, the applicability of VTC to code understanding remains insufficiently characterized.
Unlike natural language, code understanding requires strict symbol-level fidelity, and even a single missing operator or an ambiguously rendered character can break syntactic validity and compromise downstream reasoning.
It therefore remains unclear whether visual code compression can balance the trade-off between global context coverage and the high-fidelity precision required for code understanding.

To bridge the gap, we present LongCodeOCR, a visual compression framework that
renders code into compressed two-dimensional image sequences for Vision-Language Models (VLMs), and conduct a systematic empirical study on realistic SE datasets. We evaluate on a diverse suite of long-context code benchmarks spanning task types and context scales: Long Module Summarization\cite{bogomolov2024long} for assessing global semantic abstraction, LongCodeQA\cite{rando2025longcodebench} for measuring cross-file reasoning, and both Long Code Completion\cite{guo2023longcoder} and RepoBench-P\cite{bai2024longbench} for evaluating strict syntactic precision. By rendering these multi-scale contexts into images for VLM inference, we conduct an in-depth empirical examination of the visual paradigm across diverse software engineering tasks.

Through data analysis, we have summarized several insights.
First, visual code compression is more suitable for tasks that require broad semantic coverage and constraint grounding, whereas textual code compression tends to work better when the required evidence is sparse and locally salient.
Second, visual code compression exhibits more stable performance than textual code compression as context length increases, and its relative advantage becomes more pronounced for ultra-long inputs.
Third, visual code compression is governed by a fundamental coverage--fidelity trade-off: it benefits global-dependency scenarios by preserving context coverage, but can face a fidelity bottleneck on exactness-critical tasks.
In contrast, textual code compression primarily trades coverage for compression, while visual compression primarily trades symbol-level fidelity.

Our contributions can be summarized as follows:

\begin{itemize}
  \item We conduct a systematic empirical study of visual code compression for long-context code understanding.
  \item We develop LongCodeOCR, a global-preserving visual code compression framework, and evaluate it on four representative long-context benchmarks spanning code summarization, code question answering, and code completion.
  \item We analyze how visual code compression and textual code compression differ in preserving dependency-critical information, and discuss implications for designing reliable long-context code compression methods.
\end{itemize}

The rest of the paper is organized as follows: Section \ref{sec:motivation} introduces the limitations of  LongCodeZip. Section \ref{sec:framework} presents an overview of the LongCodeOCR framework. Section \ref{Study Design} describes our empirical study design. Section \ref{Study Results} records the experimental results and analyzes our findings. Section \ref{Discussion and threats to Validity} discusses the implications of our findings and potential threats to validity. Section \ref{sec:related} reviews related work concerning long-context code benchmarks, efficient long-context architectures and existing context compression paradigms. Section \ref{Conclusion} concludes the paper.

\section{Limitations of Selective Filtering in Long-Context Code Compression}
\label{sec:motivation}

LongCodeZip\cite{shi2025longcodezip} serves as a strong and representative baseline for long-context code
compression. It adopts a coarse-to-fine pipeline: it first scores and ranks function-level chunks
with Approximate Mutual Information (AMI), and then performs budget-constrained intra-function
selection via 0--1 knapsack optimization. Overall, this procedure instantiates a token-budgeted
\emph{hard selection} strategy---i.e., \emph{selective filtering-based compression}---where a
fixed budget is met through largely irreversible keep-or-discard decisions over candidate units
(e.g., functions, chunks, and blocks) guided by model-derived relevance estimates.

Although such filtering can improve performance on average, rigidly separating code into
\textbf{relevant} and \textbf{irrelevant} buckets incurs intrinsic risks. In particular, it may
fragment the available evidence and weaken the structural integrity required for reliable
reasoning: semantic dependencies often span multiple units, yet each unit can appear weakly
informative when evaluated in isolation. We next analyze these limitations along two dimensions:
\textbf{Semantic Fragmentation} (\S\ref{subsec:motivation-semantic-fragmentation}) and
\textbf{Preprocessing Overhead} (\S\ref{subsec:motivation-preprocessing-overhead}).

\subsection{How Selective Filtering Causes Semantic Fragmentation}
\label{subsec:motivation-semantic-fragmentation}

In LongCodeZip, AMI offers an actionable ranking signal by approximating how much a
candidate chunk reduces the conditional perplexity of a task-relevant target sequence. However,
this choice exposes a fundamental mismatch: perplexity-based ``contribution'' does not
necessarily capture whether a chunk is a semantic prerequisite for solving the task.

In long-context code understanding, many prerequisites take the form of \emph{dispersed and
low-salience constraints}---e.g., configuration guards, interface contracts, error-handling
conditions, cross-cutting constants, and implicit invariants. These constraints can be decisive
for solvability and correctness, yet their utility is often \emph{non-local}: their effect becomes
apparent only when multiple dependent units are available simultaneously. When candidates are
ranked at chunk granularity, such prerequisites may yield small or unstable AMI estimates and are
thus ranked behind locally predictive fragments; consequently, they are easily filtered
out.

\begin{figure}[t]
  \centering
  \includegraphics[width=\textwidth]{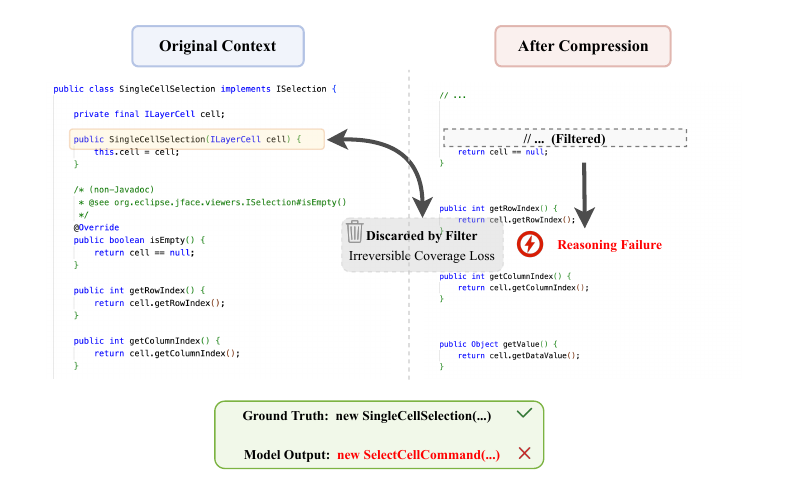}
  \caption{A motivating example from RepoBench-P illustrating how selective filtering can break dependency closure.
The \texttt{SingleCellSelection} definition (left, highlighted) is required for correct code completion but is removed after filtering (right, indicated by the masked region), causing irreversible coverage loss and an incorrect completion.}
  \label{fig:motivation}
\end{figure}

\autoref{fig:motivation} concretely illustrates this failure mode of selective filtering.
In the original context (left), the highlighted class definition of \mbox{SingleCellSelection} provides a prerequisite for correct completion, namely the constructor signature \mbox{SingleCellSelection(ILayerCell cell)}.
Although this snippet may appear weakly informative when scored in isolation, it is structurally necessary to ground the correct instantiation (bottom, ground truth).
After selective filtering (right), the definition is removed, leaving a masked region that indicates missing evidence for a valid initialization.
With the constructor information no longer available in the remaining context, the model backs off to an alternative class (\mbox{SelectCellCommand}; bottom, model output), resulting in a reasoning failure and an incorrect completion.

More broadly, source code imposes strong \emph{dependency-closure} requirements: syntax
completeness, scope and name binding, type and symbol resolution, and data-flow dependencies
constrain which evidence must co-occur to enable correct completion. \emph{Selective
filtering-based compression} deletes units irreversibly and can therefore break dependency
closure---for example, retaining a call site while dropping the callee definition.
\textbf{Once a prerequisite is removed, downstream models cannot reconstruct the missing evidence,
creating an unrecoverable recall gap that bounds achievable reliability.}

\begin{table*}[t]
  \centering
  \small
  \renewcommand{\arraystretch}{1.2}
  \setlength{\tabcolsep}{0pt}

  \caption{Compression overhead of selective filtering-based compression.
We report compression-stage latency and LLM token usage across input lengths.
Lower values indicate lower overhead, the best (lowest) values are bolded.}
Tokens are reported in millions (M).
  \label{tab:motivation-overhead}

  \begin{tabular*}{\textwidth}{@{\extracolsep{\fill}} l rrrrrrr }
    \toprule
    \textbf{Method} & \textbf{8k} & \textbf{16k} & \textbf{32k} & \textbf{64k} & \textbf{128k} & \textbf{512k} & \textbf{1M} \\
    \midrule

    \multicolumn{8}{l}{\textit{\textbf{Compression-stage Latency (seconds)}}} \\
    \hspace{1em}LongCodeZip (coarse) & 2.87 & 12.41 & 13.00 & 24.70 & 40.86 & 163.50 & 254.85 \\
    \hspace{1em}LongCodeZip (fine)   & 24.05 & 67.56 & 157.90 & 377.91 & 2774.14 & 7246.67 & 15154.77 \\
    \hspace{1em}LongCodeZip (total)  & 26.97 & 80.05 & 171.06 & 402.89 & 2815.52 & 7413.15 & 15415.53 \\
    \hspace{1em}\textbf{Visual code compression (total)}  & \textbf{0.84} & \textbf{1.52} & \textbf{2.79} & \textbf{5.30} & \textbf{9.35} & \textbf{37.20} & \textbf{70.36} \\

    \midrule

    \multicolumn{8}{l}{\textit{\textbf{Compression-stage LLM Tokens (M)}}} \\
    \hspace{1em}LongCodeZip (tokens) & 0.17 & 0.44 & 2.32 & 7.71 & 79.91 & 172.77 & 221.76 \\
    \hspace{1em}\textbf{Visual code compression (tokens)} & \textbf{0} & \textbf{0} & \textbf{0} & \textbf{0} & \textbf{0} & \textbf{0} & \textbf{0} \\

    \bottomrule
  \end{tabular*}

  \vspace{2pt}
\begin{minipage}{\textwidth}
\end{minipage}
\end{table*}

\subsection{Time and Token Cost of the Compression Stage}
\label{subsec:motivation-preprocessing-overhead}

In long-context code tasks, selective filtering-based compression typically relies on
fine-grained scoring and comparison over a large pool of candidate units.
To obtain a ranking signal suitable for \emph{selective filtering}, the compressor must run many
model forward passes (e.g., estimating conditional perplexity differences) across candidates.
\textbf{As a result, compression itself becomes a computational bottleneck, whose cost grows rapidly with context length.}

This compression-phase inference overhead directly translates into preprocessing latency,
which is difficult to reconcile with interactive workflows that require near real-time
responses.
\autoref{tab:motivation-overhead} quantifies the preprocessing burden of selective
filtering-based compression. For LongCodeZip, the coarse ranking stage grows moderately with
length (13.00 seconds at 32k to 40.86 seconds at 128k), whereas the fine-grained selection stage
dominates the compression-stage latency. At 128k, fine-grained selection takes 2774.14 seconds out of a
total of 2815.52 seconds (\textbf{98.5\%}), already pushing preprocessing to \mbox{\,$\sim$47\,min}.
At 1M, the total compression time further reaches 15415.53 seconds (\mbox{\,$\sim$4.3\,h}), with
15154.77 seconds spent in the fine-grained stage. Meanwhile, compression-stage LLM token usage
also scales sharply, from 2.32M tokens at 32k to 79.91M at 128k and 221.76M at 1M.
\textbf{These results indicate that compute pressure is not removed but shifted to the
compression phase, making it difficult to achieve both high fidelity and interactive latency.}

By contrast, our approach completes compression in 70.36 seconds at 1M without any
compression-stage LLM calls (0 tokens).
For LongCodeZip, a latency-oriented alternative is to adopt coarse-grained filtering. However, such strategies
provide limited support for preserving cross-unit dependencies and dependency closure, and thus
frequently sacrifice fidelity for speed, resulting in a characteristic \emph{efficiency--fidelity}
trade-off under ultra-long contexts.

Taken together, selective filtering-based compression can be brittle
under evidence fragmentation and can incur substantial preprocessing overhead at ultra-long
lengths, motivating alternative representations that preserve global structure.

\section{Framework Overview}
\label{sec:framework}

\begin{figure*}[t]
  \centering
  \includegraphics[width=\linewidth]{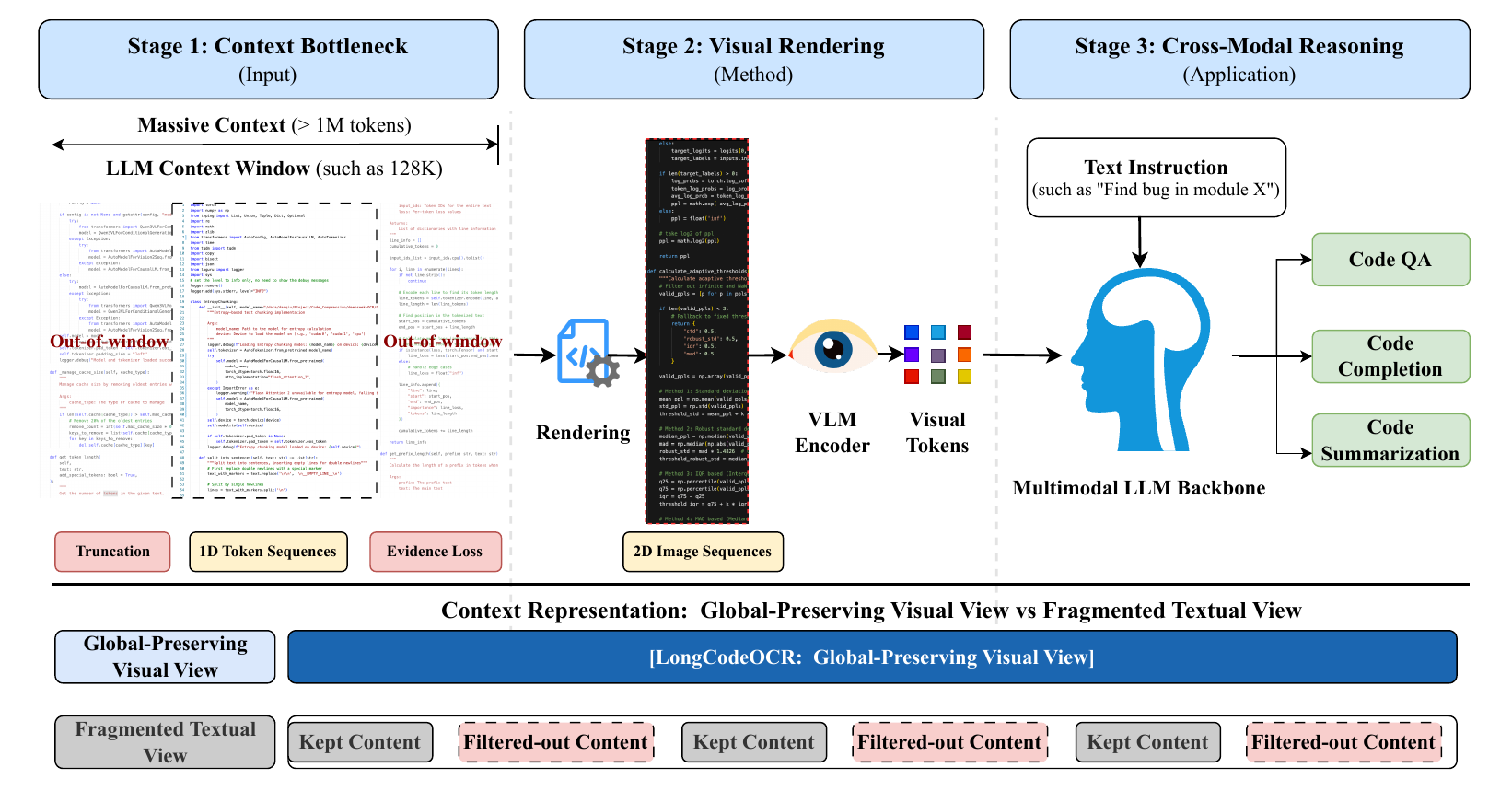}
  \caption{Framework overview of LongCodeOCR and two paradigms for long-context code compression.
  Stages~1--3 depict how massive code contexts are rendered into a 2D image sequences and encoded into visual tokens for instruction-conditioned, cross-modal reasoning.
  The bottom panel highlights the contrast between the global-preserving visual view and the fragmented textual view.}
  \label{fig:framework}
\end{figure*}

We study context compression for long-context code understanding through two paradigms: \textbf{textual code compression, which yields a fragmented textual view, and visual code compression, which maintains a global-preserving visual view.}
\autoref{fig:framework} overviews the LongCodeOCR pipeline (Stages~1--3) and and highlights the structural contrast between the fragmented textual snippets resulting from selective filtering and our unified, global-preserving visual view under the same input budget.

\paragraph{Stage 1: Context Bottleneck (Input).}
As shown in the left panel of \autoref{fig:framework}, real-world repositories can induce massive contexts that exceed the context window of standard text-only LLMs.
Under a fixed window (such as 128K tokens), code is processed as a 1D token sequences, and length control typically relies on truncation or selective filtering.
As a result, out-of-window content is omitted, which can cause evidence loss when task-critical prerequisites fall outside the retained context.

\paragraph{Stage 2: Visual Rendering (Method).}
LongCodeOCR implements global-preserving visual compression by rendering code into a compact 2D image sequences.
A vision-language encoder converts the rendered view into a sequence of visual tokens, providing a high-density representation within the same input constraint.
Unlike selective filtering on a 1D token sequences, this stage densifies context without requiring fine-grained keep-or-discard decisions over code units.

\paragraph{Stage 3: Cross-Modal Reasoning (Application).}
In the right panel of \autoref{fig:framework}, a multimodal LLM backbone performs instruction-conditioned fusion between the text instruction and the encoded visual tokens.
The model then generates task-specific outputs for downstream SE tasks, including code QA, code completion, and code summarization.

\paragraph{Context representation: global-preserving visual view vs. fragmented textual view.}
The bottom panel of \autoref{fig:framework} summarizes the contrast between the two paradigms:
\begin{itemize}
  \item \textbf{Textual Code Compression.}
  This paradigm meets the input constraint via keep-or-discard decisions over candidate units, producing kept content interleaved with filtered-out content.
  Because semantic dependencies can span multiple units, selective filtering may remove task-critical prerequisites, weakening dependency closure and reducing reliability.

  \item \textbf{Visual Code Compression.}
  In contrast, LongCodeOCR renders the context into a global-preserving view and encodes it as visual tokens, yielding a more contiguous representation of global structure under the same constraint.
  This approach avoids the dependency breakage inherent in selective filtering by preserving a global view of the context.
\end{itemize}

\textbf{In this study, we instantiate the visual code compression paradigm through a proposed framework, LongCodeOCR. We use LongCodeOCR as the primary vehicle to empirically evaluate the efficacy of the visual paradigm against textual code compression methods.}

\section{Study Design}
\label{Study Design}

\subsection{Datasets}
We study three representative long-context code understanding tasks: code summarization for assessing global semantic abstraction, code question answering for measuring cross-file reasoning, and code completion for evaluating strict syntactic precision.
Collectively, these tasks evaluate whether a compressed context preserves task-critical evidence for reliable performance in developer-facing scenarios.

For the code summarization task, we adopt the Long Module Summarization dataset\cite{bogomolov2024long}, which targets project-level documentation generation given a project code context and a one-sentence \emph{intent}.
To target challenging long-context settings, we retain only 139 instances whose contexts exceed 2{,}000 tokens.

For the code question answering task, we adopt LongCodeQA\cite{rando2025longcodebench}, a four-option multiple-choice benchmark for evaluating long-context code understanding and reasoning.
Each instance includes an instruction, a filename-sorted set of repository files, and a question derived from resolved issues in public GitHub Python repositories via an automated pipeline; questions that remain answerable without repository context are removed.
The benchmark contains 443 instances from 98 repositories and is balanced across context-length ranges.

For the code completion task, we evaluate it across two distinct scopes: \textbf{file-level}, which focuses on internal consistency, and \textbf{repository-level}, which demands cross-file reasoning.

\begin{itemize}
    \item \textbf{Long Code Completion (File-level context).} 
    Introduced in LongCoder~\cite{guo2023longcoder}, Long Code Completion(LCC) targets completion within a single file. 
    It evaluates the model's capacity to maintain local semantic consistency while capturing \textit{long-range constraints} appearing earlier in the same file. 
    To emphasize long-context difficulty, we restrict the test set to 500 Python instances whose input contexts exceed 5{,}000 tokens.
    
    \item \textbf{RepoBench-P (Repository-level context).} 
    Adapted from LongBench~\cite{bai2024longbench}, this benchmark targets completion requiring \textit{cross-file evidence}. 
    Unlike LCC, it constructs inputs by retrieving relevant fragments from external files (e.g., via import statements). 
    We specifically employ the \textbf{XF-F (Cross-File-First)} setting, where within-file context alone is insufficient, thereby directly assessing the model's ability to integrate and leverage distributed dependencies.
\end{itemize}

\subsection{Model and Baseline Selection}
\subsubsection{Baseline Selection}
\label{sec:baseline-selection}

We evaluate LongCodeOCR against a variety of competitive baselines:

\begin{itemize}
  \item \textbf{No Compression:} We provide the full code context without applying any compression.

  \item \textbf{Random Baseline:} Random Line removes complete lines of code uniformly at random.

  \item \textbf{Retrieval-based Methods:} We instantiate two RAG variants: RAG (Sliding Window) segments the context into fixed-size overlapping chunks, whereas RAG (Function Chunking) splits code at function boundaries. Both variants represent chunks with UniXCoder-base and select the top-$k$ chunks by instruction--chunk similarity for concatenation.\cite{guo2022unixcoder,zhang2023repocoder}

  \item \textbf{Text Compression Methods:} We apply representative prompt-compression methods to the code context, including LLMLingua~\cite{jiang2023llmlingua}, LongLLMLingua~\cite{jiang2024longllmlingua}, and LLMLingua-2\cite{pan2024llmlingua}.

  \item \textbf{Code Compression Methods:} We include LongCodeZip~\cite{shi2025longcodezip} as a strong code-aware compression baseline. It implements a coarse-to-fine pruning strategy that filters context by ranking function-level chunks via approximate mutual information and performing fine-grained block selection to retain instruction-relevant spans.
\end{itemize}

\subsubsection{Model Selection}
\label{sec:model-selection}

We evaluate open-source models of similar parameter scales to isolate the effect of context representation on long-context code tasks under a consistent evaluation protocol.
In particular, we compare the same code context represented as plain text versus as rendered images.
Specifically, we use Qwen3-8B\cite{yang2025qwen3technicalreport} as the text-only LLM baseline, and include two vision--language models (VLMs) as multimodal counterparts: Qwen3-VL-8B\cite{bai2025qwen3vltechnicalreport} and Glyph\cite{cheng2025glyph} (9B).
Qwen3-VL-8B serves as a general-purpose VLM baseline for direct visual code understanding with off-the-shelf multimodal capabilities.
In contrast, Glyph is a long-context-oriented VLM designed for visual compression: it employs an LLM-guided search to tune rendering configurations and is trained with a dedicated vision--text alignment curriculum.
Together, these components enable Glyph to encode substantially more textual evidence into a compact visual token budget than a general-purpose VLM.
Although Glyph is slightly larger than the 8B baselines, they remain within a comparable parameter regime, enabling a fair contrast between general-purpose and specialized visual compression models under similar capacity constraints.
To ensure rigorous comparison, we employ identical prompt templates and consistent evaluation protocols across all tasks.

\subsection{Evaluation Metrics}
\subsubsection{Compression Ratio.}

We evaluate compression ratios across two input modalities: (i) visual code compression, where contexts are rendered and encoded as visual tokens (represented by LongCodeOCR), and (ii) textual code compression, where contexts are reduced via  pruning or filtering (represented by LongCodeZip and other baselines).

\textbf{Visual code compression.} 
For the visual code compression method, let $C_{\text{context}}$ denote the \textit{token sequence} of the original code context, counted using the tokenizer of the evaluated VLM. 
After rendering, the code context is represented as a sequence of $n$ images $\mathcal{V}=\{v_i\}_{i=1}^{n}$.

To quantify the information cost, we adopt the estimation formula from VTCBench~\cite{zhao2025vtcbench}. 
We abstract away model-specific resizing and assume each page $v_i$ has resolution $(H_i, W_i)$. 
The \textit{estimated} number of visual tokens $\tau(v_i)$ consumed by the VLM is calculated as:
\begin{equation}
\tau(v_i)=
\begin{cases}
\frac{H_i \cdot W_i}{16 \cdot 16 \cdot 4}, & \text{Qwen3-VL-8B}~\cite{bai2025qwen3vltechnicalreport}\\
\frac{H_i \cdot W_i}{14 \cdot 14 \cdot 4}, & \text{Glyph}~\cite{cheng2025glyph}
\end{cases}
\end{equation}
\noindent where the denominator factors account for the patch size ($16^2$ or $14^2$) and the $2 \times 2$ pooling mechanism (factor $4$) inherent to the VLM architectures.

The total visual token count is $|C_{\text{visual}}| = \sum_{i=1}^{n} \tau(v_i)$.
The compression ratio is computed as:
\begin{equation}
\mathrm{Ratio}_{\text{visual}} = \frac{\lvert C_{\text{context}} \rvert}{\lvert C_{\text{visual}} \rvert}
\end{equation}
\noindent where $\lvert C_{\text{context}}\rvert$ and $\lvert C_{\text{visual}}\rvert$ denote the total token counts of the original code context and the rendered visual context, respectively.

\textbf{Textual code compression.} 
For the textual code compression methods, the ratio is defined as:
\begin{equation}
\mathrm{Ratio}_{\text{textual}} = \frac{\left|C_{\text{context}}\right|}{\left|C_{\text{compressed}}\right|}
\end{equation}
\noindent where $\lvert C_{\text{compressed}}\rvert$ denotes the token count of the compressed context.

\subsubsection{Downstream Task Performance.}
We evaluate downstream performance using task-specific metrics.

For the code summarization task, we follow~\cite{bogomolov2024long} and use GPT-4o-mini~\cite{openai2024gpt4ocard} as a third-party referee model to perform pairwise comparison between the generated summary $s_o$ and the reference summary $\hat{s}$, conditioning on the code.
To mitigate order bias, we query the referee twice with reversed order.
We then compute \emph{CompScore} as:
\begin{equation}
\mathrm{CompScore} = \frac{1}{2}\Big[ P(s_o \succ \hat{s}) + \big(1 - P(\hat{s} \succ s_o)\big) \Big],
\end{equation}
where $P(s_o \succ \hat{s})$ is the probability that the referee prefers $s_o$ over $\hat{s}$, and $P(\hat{s} \succ s_o)$ is the probability under the reversed order.
$\mathrm{CompScore}$ ranges from 0 to 100, with 50 indicating equal preference.

For the code completion task, we follow benchmark-specific evaluation protocols: for LCC, we follow LongCoder~\cite{guo2023longcoder} to evaluate model performance using Exact Match (EM) and Edit Similarity (ES). For RepoBench-P, we follow LongBench~\cite{bai2024longbench} to evaluate model performance using ES.

For the code question answering task, we follow LongCodeBench~\cite{rando2025longcodebench} to report answer accuracy.

\subsection{Implementation Details}

\begin{table}[t]
\centering
\caption{Rendered context fields and benchmark-level average compression ratios.}
\label{tab:rendering-config}
\small
\renewcommand{\arraystretch}{1.15}
\setlength{\tabcolsep}{6pt}
\begin{tabular}{l l l r}
\toprule
Task & Dataset & Rendered context field & Avg. ratio \\
\midrule
Code Summarization & Long Module Summarization & \texttt{context} & 1.7 \\
Code Question Answering & LongCodeQA & \texttt{repo\_text} & 1.6 \\
\multirow{2}{*}{Code Completion} & Long Code Completion & \texttt{background\_context} & 2.0 \\
 & RepoBench-P & \texttt{context} & 2.0 \\
\bottomrule
\end{tabular}
\end{table}

To avoid introducing answer-related information during visualization and to ensure a fair comparison, we render only the long code context that serves as model input into images for VLM-based methods. Non-code inputs, such as task instructions, intents, and candidate options, are kept in text form. LLM baselines and VLM-based methods are provided with the same information; they differ only in how the long context is represented (text vs. images).

For the LCC task, the function to be completed appears at the end of the sample context. Rendering the entire context may expose the target code and thus cause information leakage. Following the processing protocol of LongCodeZip~\cite{shi2025longcodezip}, we split the target function from the original sample and render only the preceding code snippet as the visualized context (i.e., \texttt{background\_context}). The model is  tasked with completing the separated target function implementation given this context.

Table~\ref{tab:rendering-config} summarizes, for each benchmark, the rendered field used by VLM-based methods and the corresponding average compression ratio. To ensure comparability, we keep the settings in Table~\ref{tab:rendering-config} fixed within each benchmark and evaluate all methods under the same compression ratio specified in the table. Therefore, performance differences across methods cannot be attributed to different compression ratios, but rather reflect the effectiveness of the methods themselves.

Regarding inference, explicit reasoning mode, denoted as ``thinking'', is enabled for all evaluated models. The reasoning budget is capped at 2048 tokens; once the cap is reached, the reasoning phase is terminated and the final answer is produced. For all tasks, evaluation is conducted solely on the final outputs, while intermediate reasoning content is excluded. Except for enabling ``thinking'' and setting the associated budget, decoding parameters and output constraints are kept identical across models to reduce confounding effects introduced by generation strategies.

\section{Study Results}
\label{Study Results}
In this section, we present experimental results and our analysis to answer the research questions.

\subsection{RQ1: Can visual code compression serve as a viable alternative to textual code compression for long-context understanding?}
\label{RQ1}

\begin{table}[t]
\centering
\caption{CompScore on Long Module Summarization at average compression ratio $\approx$1.7$\times$. Best results are bolded.}
\label{tab:rq1-lms}
\begin{tabular*}{\linewidth}{@{\extracolsep{\fill}}llcc@{}}
\toprule
\textbf{Model} & \textbf{Method} & \textbf{CompScore} & \textbf{Ratio} \\
\midrule
\multirow{8}{*}{Qwen3-8B}
  & \textit{No Compression}   & 47.55 & 1.0$\times$ \\
  & Random Line               & 52.07 & 1.7$\times$ \\
  & RAG (Sliding Window)      & 43.05 & 1.7$\times$ \\
  & RAG (Function Chunking)   & 46.34 & 1.7$\times$ \\
  & LongLLMLingua             & 45.76 & 1.7$\times$ \\
  & LLMLingua                 & 48.23 & 1.8$\times$ \\
  & LLMLingua-2               & 58.18 & 1.7$\times$ \\
  & LongCodeZip               & 50.40 & 1.7$\times$ \\
\midrule
Qwen3-VL-8B & \textbf{LongCodeOCR (ours)} & \textbf{87.25} & \textbf{1.7$\times$} \\
\bottomrule
\end{tabular*}
\end{table}

To assess whether visual code compression can serve as a reliable alternative to textual representations without sacrificing downstream utility, we evaluate our instantiation, LongCodeOCR, on the Long Module Summarization benchmark introduced by LongCodeZip~\cite{shi2025longcodezip}.
We adopt this benchmark for three reasons.
First, summarization over long repositories is highly sensitive to missing prerequisites and non-local dependencies, making it a stringent testbed for context compression.
Second, it provides a standardized evaluation protocol and a broad suite of representative baselines.
Third, LongCodeZip reports its most competitive results on this benchmark among the tasks in their study; using the same setting therefore yields a direct and baseline-favorable comparison.

Table~\ref{tab:rq1-lms} reports CompScore together with the achieved average compression ratio for each method.
We focus on the target compression strength with a ratio of approximately 1.7$\times$.
Non-visual baselines operate on code tokens and are evaluated with Qwen3-8B.
LongCodeOCR renders code as images and is evaluated with a VLM, Qwen3-VL-8B.
Overall, LongCodeOCR achieves a substantially higher CompScore than all non-visual baselines at comparable compression ratios, supporting the effectiveness of global-preserving visual code compression under strong compression.

Within non-visual baselines, moderate compression is not necessarily detrimental.
Random line sampling at ratio 1.7$\times$ improves CompScore from 47.55 under no compression to 52.07, indicating redundancy in long-context inputs.
Retrieval-based variants are less effective in this setting.
Sliding-window retrieval achieves 43.05, and function-level chunking reaches 46.34, both below the no-compression baseline.
Prompt compression methods exhibit mixed results.
LongLLMLingua and LLMLingua obtain 45.76 and 48.23, respectively.
LLMLingua-2 yields the strongest performance among non-visual baselines and reaches 58.18.
Also, LongCodeZip achieves 50.40 at the target ratio.

LongCodeOCR shows a large advantage at the same compression strength.
At ratio 1.7$\times$, LongCodeOCR reaches a CompScore of 87.25 and exceeds LLMLingua-2 by 29.07 points.
This margin is far larger than the gain from simple redundancy removal, where random line sampling improves by 4.52 points over no compression.
These results suggest that representing code in a 2D visual layout can retain task-relevant signals more effectively than keep-or-discard selection on 1D token sequences.
In particular, rendering preserves layout cues such as indentation, alignment, and block structure.
It also provides a more contiguous view of code regions that may otherwise be fragmented by selective filtering or retrieval.

\begin{findingbox}
\textbf{Finding 1:} Visual code compression significantly outperforms textual code compression on code summarization at comparable ratios ($\approx$1.7$\times$), demonstrating superior preservation of global semantic integrity
\end{findingbox}

\subsection{RQ2: How does visual code compression compare with textual code compression across tasks and context lengths?}
\label{RQ2}

\begin{table}[t]
  \centering
  \small
  \setlength{\tabcolsep}{5pt}
  \renewcommand{\arraystretch}{1.15}
  \caption{Visual code compression versus textual code compression on Long Module Summarization, stratified by input length. Each entry reports CompScore. Compression ratios are approximately comparable ($\approx$1.7$\times$). Best results are bolded. $n$ denotes the number of instances in each length bin.}
  \label{tab:rq2-summary}

  \begin{tabular}{@{}ll ccc cc@{}}
    \toprule
    \textbf{Method} & \textbf{Model} 
      & \multicolumn{3}{c}{\textbf{Input length}} 
      & \textbf{Avg} 
      & \textbf{Ratio} \\
    \cmidrule(lr){3-5}
     &  
      & \footnotesize 0k--8k (n=64) 
      & \footnotesize 8k--16k (n=31) 
      & \footnotesize 16k+ (n=44) 
      &  &  \\
    \midrule

    LongCodeZip  &  Qwen3-8B & 56.47 & 52.56 & 40.06 & 50.40 & 1.7$\times$ \\
    
    \midrule
    
    \multirow{2}{*}{LongCodeOCR} 
      & Qwen3-VL-8B & \textbf{89.09} & \textbf{92.65} & \textbf{80.76} & \textbf{87.25} & 1.7$\times$ \\
      & Glyph       & 73.49 & 74.29 & 70.56 & 72.74 & 1.7$\times$ \\

    \bottomrule
  \end{tabular}
\end{table}

To examine how code compression strategies generalize across tasks and increasing context lengths, we compare LongCodeOCR with a representative textual code compression,  LongCodeZip~\cite{shi2025longcodezip} on a suite of long-context code benchmarks. For each task, we stratify evaluation instances by input length and report task-specific performance within each length stratum. We additionally include Glyph\cite{cheng2025glyph}, a VLM) tailored for long-context visual compression, to assess whether LongCodeOCR is effectiveness or robust across both general-purpose and specialized VLMs.

\begin{table}[h]
  \centering
  \small
  \setlength{\tabcolsep}{5pt}
  \renewcommand{\arraystretch}{1.15}
  \caption{Visual code compression versus textual code compression on RepoBench-P, stratified by input length. Each entry reports ES. Compression ratios are approximately comparable ($\approx$2.0$\times$). Best results are bolded. $n$ denotes the number of instances in each length bin.}
  \label{tab:rq2-repobench}

  \begin{tabular}{@{}ll ccc cc@{}}
    \toprule
    \textbf{Method} & \textbf{Model} 
      & \multicolumn{3}{c}{\textbf{Input length}} 
      & \textbf{Avg} 
      & \textbf{Ratio} \\
    \cmidrule(lr){3-5}
     &  
      & \footnotesize 0k--4k (n=278) 
      & \footnotesize 4k--8k (n=180) 
      & \footnotesize 8k+ (n=42) 
      &  &  \\
    \midrule

    LongCodeZip & Qwen3-8B & 42.25 & 37.37 & 30.17 & 39.48 & 2.0$\times$ \\
    
    \midrule
    
    \multirow{2}{*}{LongCodeOCR} 
      &  Qwen3-VL-8B & 52.33 & 47.98 & 45.76 & 50.22 & 2.0$\times$ \\
      & Glyph       & \textbf{62.20} & \textbf{59.88} & \textbf{53.98} & \textbf{60.68} & 2.0$\times$ \\

    \bottomrule
  \end{tabular}
\end{table}

\subsubsection{Performance Analysis across Diverse Code Task Types}
We compare two compression \emph{approaches}, visual code compression via LongCodeOCR and textual code compression exemplified by LongCodeZip, using the average (AVG) results across benchmarks in Tables~\ref{tab:rq2-summary}--\ref{tab:rq2-lqa-ultra}.

\textbf{Code summarization: Global abstraction with long-range dependencies.} 
Table~\ref{tab:rq2-summary} reveals a substantial advantage for visual compression. 
Under aligned budgets ($\approx 1.7\times$), LongCodeOCR with Qwen3-VL-8B dominates LongCodeZip (50.40) with a CompScore of \textbf{87.25} (\textbf{+36.85}). 
This intrinsic benefit is corroborated by the specialized Glyph (72.74). 
We attribute these gains to the preservation of global spatial layout, whereas textual code compression risks fragmenting context by discarding globally informative cues. 
The gap between VLMs further indicates that summarization relies on code-aware semantic organization.

\textbf{Code completion: Constraint recovery and symbolic precision.}
Under an approximately comparable compression ratio of $2.0\times$, visual code compression consistently improves code completion over textual code compression.
On RepoBench-P (Table~\ref{tab:rq2-repobench}), the average ES increases from 39.48 with LongCodeZip to 50.22 when LongCodeOCR uses Qwen3-VL-8B as the backbone, and further to 60.68 when it uses Glyph; on LCC (Table~\ref{tab:rq2-lcc}), the average ES/EM increases from 36.93/10.60 to 42.21/12.00 and 45.90/13.60.
Mechanistically, discard-based textual code compression incurs irreversible coverage loss that is more damaging when constraints are dispersed across files, whereas visual code compression preserves broader coverage but is bounded by fidelity for fine-grained symbols; a code-specialized visual backbone (Glyph) further mitigates this fidelity bottleneck under exactness requirements.

\begin{table}[t]
  \centering
  \small
  \setlength{\tabcolsep}{5pt}
  \renewcommand{\arraystretch}{1.15}
  \caption{Visual code compression versus textual code compression on LCC, stratified by input length. Each entry reports ES/EM (left/right). Compression ratios are approximately comparable ($\approx$2.0$\times$). Best results are bolded. $n$ denotes the number of instances in each length bin.}
  \label{tab:rq2-lcc}

  \begin{tabular*}{\columnwidth}{@{\extracolsep{\fill}}ll ccc cc@{}}
    \toprule
    \textbf{Method} & \textbf{Model}
      & \multicolumn{3}{c}{\textbf{Input length}}
      & \textbf{Avg}
      & \textbf{Ratio} \\
    \cmidrule(lr){3-5}
     &
      & \footnotesize 0k--8k (n=275)
      & \footnotesize 8k--16k (n=177)
      & \footnotesize 16k+ (n=48)
      &  &  \\
    \midrule

   LongCodeZip & \shortstack[l]{ Qwen3-8B}
      & 38.48\,/\,9.45
      & 35.14\,/\,12.99
      & 34.68\,/\,8.33
      & 36.93\,/\,10.60
      & 2.0$\times$ \\

    \midrule

    \multirow{2}{*}{LongCodeOCR}
      & \shortstack[l]{Qwen3-VL-8B}
      & 42.18\,/\,13.45
      & 43.28\,/\,12.43
      & 38.38\,/\,2.08
      & 42.21\,/\,12.00
      & 2.0$\times$ \\

      & \shortstack[l]{Glyph}
      & \textbf{45.56}\,/\,\textbf{13.45}
      & \textbf{46.91}\,/\,\textbf{13.56}
      & \textbf{44.09}\,/\,\textbf{14.58}
      & \textbf{45.90}\,/\,\textbf{13.60}
      & 2.0$\times$ \\

    \bottomrule
  \end{tabular*}
\end{table}

\textbf{Code QA: Evidence-seeking and reasoning over long contexts.} 
Tables~\ref{tab:rq2-lqa} and \ref{tab:rq2-lqa-ultra} reveal a regime-dependent competitive landscape where neither paradigm uniformly dominates. 
In the standard long-context regime (Table~\ref{tab:rq2-lqa}), LongCodeOCR (Qwen3-VL-8B) maintains a slight edge over LongCodeZip (70.46\% vs.\ 67.61\%), while the specialized model Glyph trails at 64.42\%. 
However, the ultra-long regime (Table~\ref{tab:rq2-lqa-ultra}) exposes the superior scalability of visual compression.

While LongCodeZip leads at 512k, a decisive crossover occurs at 1M. 
Strikingly, Glyph achieves \textbf{70.00\%} accuracy under an extreme compression ratio of \textbf{12.2$\times$}, significantly outperforming LongCodeZip (64.00\%) which operates at a much lower compression of 3.0$\times$. 
Along with Qwen3-VL-8B (72.00\%), these results suggest that as context scales to millions of tokens, preserving global visual layout offers higher information density than text filtering, enabling effective evidence retrieval even under severe compression constraints.

\begin{table}[h]
  \centering
  \small
  \setlength{\tabcolsep}{5pt}
  \renewcommand{\arraystretch}{1.15}
 \caption{Visual code compression versus textual code compression on LongCodeQA, stratified by input length. Each entry reports accuracy (Acc\%). Compression ratios are approximately comparable ($\approx$1.6$\times$). Best results are bolded. $n$ denotes the number of instances in each length bin.}
  \label{tab:rq2-lqa}

  \begin{tabular}{@{}ll ccc cc@{}}
    \toprule
    \textbf{Method} & \textbf{Model} 
      & \multicolumn{3}{c}{\textbf{Input length}} 
      & \textbf{Avg} 
      & \textbf{Ratio} \\
    \cmidrule(lr){3-5}
     &  
      & \footnotesize 32k (n=113) 
      & \footnotesize 64k (n=76) 
      & \footnotesize 128k (n=92) 
      &  &  \\
    \midrule

    LongCodeZip &  Qwen3-8B & \textbf{64.60} & 71.05 & 68.48 & 67.61 & 1.6$\times$ \\
    
    \midrule
    
    \multirow{2}{*}{LongCodeOCR} 
      & Qwen3-VL-8B & 63.72 & \textbf{80.26} & \textbf{70.65} & \textbf{70.46} & 1.6$\times$ \\
      & Glyph       & 63.72 & 72.37 & 58.70 & 64.42 & 1.6$\times$ \\

    \bottomrule
  \end{tabular}
\end{table}

\begin{table}[t]
  \centering
  \small
  \setlength{\tabcolsep}{4.5pt}
  \renewcommand{\arraystretch}{1.15}
  \caption{Scalability at ultra-long inputs (512k--1M) on LongCodeQA. Each entry reports accuracy (Acc\%). Compression ratios vary due to model context limits. Best results are bolded. $n$ denotes the number of instances in each length bin.}
  \label{tab:rq2-lqa-ultra}

  \begin{tabular}{@{}ll rr rr@{}}
    \toprule
    \textbf{Method} & \textbf{Model}
      & \multicolumn{2}{c}{\textbf{512k (n=47)}}
      & \multicolumn{2}{c}{\textbf{1M (n=50)}} \\
    \cmidrule(lr){3-4} \cmidrule(lr){5-6}
     &  & \textbf{Acc\%} & \textbf{Ratio} & \textbf{Acc\%} & \textbf{Ratio} \\
    \midrule

    LongCodeZip & Qwen3-8B
      & \textbf{74.47} & 2.8$\times$ & 64.00 & 3.0$\times$ \\

    \midrule

    \multirow{2}{*}{LongCodeOCR} & Qwen3-VL-8B
      & 65.96 & 2.8$\times$ & \textbf{72.00} & 8.1$\times$ \\
     &  Glyph
      & 59.57 & 3.7$\times$ & 70.00 & 12.2$\times$ \\

    \bottomrule
  \end{tabular}
\end{table}

\begin{findingbox}
\textbf{Finding 2}: Visual code compression excels in tasks requiring semantic coverage and constraint grounding, whereas textual code compression is effective for tasks driven by sparse, locally salient evidence.
\end{findingbox}

\subsubsection{Performance Analysis across Varying Code Context Length}
Stratified evaluations across benchmarks indicate that visual code compression is generally more robust to context scaling than selective filtering, while also exposing task-dependent bottlenecks at extreme lengths.

\textbf{Robustness against context expansion. }
For semantic-heavy tasks, including summarization and repo-level code completion, a consistent pattern emerges as inputs extend to the longest bins: LongCodeZip degrades markedly, whereas LongCodeOCR retains stronger utility.
In summarization (Table \ref{tab:rq2-summary}), the textual baseline drops from 56.47 (0k--8k) to 40.06 (16k+), while LongCodeOCR with Qwen3-VL-8B remains at 80.76.
A similar trend appears in repo-level code completion (Table~\ref{tab:rq2-repobench}), where LongCodeZip decreases from 42.25 to 30.17, in contrast to the visual model’s more stable performance (45.76).
These divergences are consistent with different degradation mechanisms: textual code compression can induce \textbf{context fragmentation} by pruning dispersed yet decisive cues, whereas visual code compression mitigates coverage loss by retaining more global structural information.

\textbf{Precision bottlenecks at extreme lengths. }
The exactness-sensitive LCC task (Table~\ref{tab:rq2-lcc}) further highlights a fidelity boundary of visual encoding under high-density inputs.
While Glyph remains comparatively stable, Qwen3-VL-8B exhibits a sharp drop in Exact Match in the 16k+ regime (EM: 2.08) despite maintaining moderate semantic similarity (ES: 38.38).
This pattern is consistent with a \textbf{visual crowding} effect: as context density increases, glyphs become smaller and layouts more crowded, amplifying symbol-level noise.
Such noise may have limited impact on coarse semantic similarity, but it becomes a primary bottleneck when strict symbol-level fidelity is required.

\textbf{Scalability and budget-sensitive reversal.}
Finally, the LongCodeQA benchmark (Tables~\ref{tab:rq2-lqa}--\ref{tab:rq2-lqa-ultra}) illustrates a regime-dependent shift under feasible budgets.
In the standard long-context regime (Table~\ref{tab:rq2-lqa}), LongCodeOCR attains strong performance at moderate lengths (e.g., 80.26 vs.\ 71.05 at 64k).
More notably, the ultra-long regime (Table~\ref{tab:rq2-lqa-ultra}) reveals a crossover: LongCodeZip leads at 512k (74.47), whereas LongCodeOCR becomes more competitive at 1M (72.00 vs.\ 64.00).
We interpret this reversal as utility achieved under \textbf{fit-in-context budgets}: at million-token scales, where feasibility is constrained by context limits, visual compression can offer higher effective information density than textual code compression.

\begin{table}[h]
  \centering
  \small
  \setlength{\tabcolsep}{5pt}
  \renewcommand{\arraystretch}{1.15}
  \caption{Reference comparison with selected long-context LLMs on LongCodeQA (Acc\%) across input lengths.  Our result uses LongCodeOCR to compress the input before inference with Qwen3-VL-8B (ratio-matched setting as in RQ2; no task-specific tuning). Best results are bolded. $n$ denotes the number of instances in each length bin.}
  \label{tab:lqa_ref_subset}

  \begin{tabular*}{\columnwidth}{@{\extracolsep{\fill}}lccccc@{}}
    \toprule
    \textbf{Model} & \textbf{32K (n=113)} & \textbf{64k (n=76)} & \textbf{128k (n=92)} & \textbf{512k (n=47)} & \textbf{1M (n=50)} \\
    \midrule
    Llama 3.1 -- 405B Instruct  & 69.9  & 72.4  & 67.4  & --    & --    \\
    GPT-4o                     & 65.5  & 76.3  & \textbf{74.3} & --    & --    \\
    Gemini 2.5 Pro             & \textbf{75.2}  & 71.1  & 71.7  & \textbf{68.1}  & 69.8  \\
    Claude 3.5 Sonnet          & 65.5  & 69.7  & 71.7  & --    & --    \\
    Claude 3.5 Sonnet + RAG    & 25.55 & 31.18 & 24.83 & 17.19 & 12.77 \\
    \midrule
    LongCodeOCR + Qwen3-VL-8B & 63.72 & \textbf{80.26} & 70.65 & 65.96 & \textbf{72.00} \\
    \bottomrule
  \end{tabular*}
\end{table}

\textbf{Reference comparison with strong open- and closed-source models.}
Table~\ref{tab:lqa_ref_subset} provides a reference comparison with representative open- and closed-source models evaluated across input lengths, where these baselines rely on their native context windows.
Without task-specific fine-tuning and with compression enabled only at inference time, LongCodeOCR with a lightweight Qwen3-VL-8B backbone achieves competitive accuracy in certain length regimes.
Specifically, at 64k input length, it reaches \textbf{80.26}, exceeding GPT-4o (76.3) and Llama~3.1-405B Instruct (72.4); at 1M input length, it attains \textbf{72.00}, slightly higher than Gemini~2.5~Pro (69.8).
These results suggest that, under feasible budget and context constraints, long-code understanding is influenced not only by native context length but also by how the context is represented; visual code compression can serve as a compact representation that improves the utility of ultra-long inputs for evidence seeking and reasoning.

\begin{findingbox}
\textbf{Finding 3:} Visual code compression scales more stably than textual code compression as context length grows, and its advantage becomes more pronounced for ultra-long inputs.
\end{findingbox}

\subsection{RQ3: What is the sensitivity of visual code compression to variations in compression ratio and rendering configuration?}

We conduct a sensitivity analysis of visual code compression on RepoBench-P\cite{bai2024longbench} and LCC\cite{guo2023longcoder} by varying compression ratio and rendering configuration.
We evaluate LongCodeOCR with Qwen3-VL-8B and compare against a textual code compression mwthod implemented by LongCodeZip with Qwen3-8B.
Due to the high overhead of fine filtering (Table~\ref{tab:motivation-overhead}), we use coarse filtering only for LongCodeZip.

\begin{table}[h]
\caption{Sensitivity of ES to compression ratio on RepoBench and LCC (4k--8k, 180 instances per benchmark). LongCodeZip uses coarse filtering only,  Row-wise maxima are bolded.}
\label{tab:rq3_ratio_sensitivity}
\centering
\small
\setlength{\tabcolsep}{4pt}
\renewcommand{\arraystretch}{1.08}
\begin{tabular}{llccccc}
\toprule
Task & Method & 1.0$\times$ & 2.0$\times$ & 3.0$\times$ & 4.0$\times$ & 5.0$\times$ \\
\midrule
\multirow{2}{*}{RepoBench-P} & LongCodeZip (coarse) & 32.75 & 40.86 & 40.44 & \textbf{45.60} & 45.36 \\
 & LongCodeOCR & 49.56 & 47.98 & 48.48 & \textbf{51.64} & 50.70 \\
\midrule
\multirow{2}{*}{LCC} & LongCodeZip (coarse) & 34.03 & \textbf{35.74} & 35.24 & 34.14 & 33.88 \\
 & LongCodeOCR & \textbf{42.83} & 39.35 & 38.95 & 35.52 & 35.34 \\
\bottomrule
\end{tabular}
\end{table}

\paragraph{Compression-ratio sensitivity.}
In terms of performance, LongCodeOCR exhibits remarkable stability on the repository-level task (RepoBench-P), with its Edit Similarity (ES) fluctuating within a narrow range of 3.66 points (47.98--51.64). In contrast, LongCodeZip shows high volatility with a range of 12.85 points, allowing the visual paradigm to maintain a significant lead across compression ratios from $1\times$ to $5\times$. However, on the single-file task (LCC), LongCodeOCR displays a monotonic degradation as compression intensifies, with a total decrease of 7.49 points, whereas the filtering-based baseline remains relatively flat with a range of only 1.86. Consequently, the performance gap narrows sharply from 8.80 at $1\times$ to 1.46 at $5\times$, indicating that increasing compression ratios exposes distinct performance bottlenecks for each paradigm.

\begin{table}
\setlength{\intextsep}{2pt}
\setlength{\textfloatsep}{4pt}
\caption{Exploratory analysis of rendering settings. We vary font and layout (single column vs. two column) while keeping all other settings fixed. We report ES on RepoBench-P and both ES/EM on LCC.}
\label{tab:rq3_render_ablation}
\centering
\small
\setlength{\tabcolsep}{4pt}
\renewcommand{\arraystretch}{1.06}
\begin{tabular}{lccc}
\toprule
Setting & RepoBench-P & \multicolumn{2}{c}{LCC} \\
\cmidrule(lr){3-4}
 & ES & ES & EM \\
\midrule
\multicolumn{4}{@{}l}{\emph{(a) Font}} \\
Verdana & 51.63 & 39.35 & 11.11 \\
JetBrainsMono-Regular & 50.98 & 41.75 & 12.78 \\
\addlinespace[2pt]
\midrule
\multicolumn{4}{@{}l}{\emph{(b) Layout}} \\
Single olumn & 51.63 & 39.35 & 11.11 \\
Two column & 49.19 & 39.77 & 10.00 \\
\bottomrule
\end{tabular}
\end{table}

 Mechanistically, this divergence stems from the fundamental difference in the ``cost'' each paradigm pays for compression. Textual code compression primarily increases the ratio by discarding code units, resulting in irreversible coverage loss. This coverage fragility is particularly detrimental to repository-level tasks where constraints are dispersed and missing a single prerequisite can be fatal; visual code compression mitigates this risk by preserving the global context. Conversely, visual code compression achieves its target by packing nearly the full context into a constrained visual budget, shifting the bottleneck toward visual fidelity. Smaller glyphs and rendering noise impair the resolution of local details required for LCC tasks, thereby diminishing the advantage of visual compression at extreme ratios. This \textbf{coverage--fidelity trade-off} explains the sustained superiority of the visual paradigm in broad reasoning scenarios and its potential limitations in local symbolic precision.

\paragraph{Rendering sensitivity.}
LongCodeOCR exhibits task-specific sensitivity to rendering configurations. In RepoBench-P, font choice has a negligible impact (51.63 vs. 50.98), but two-column layouts reduce ES from 51.63 to 49.19. Conversely, LCC is more precision-dependent: JetBrainsMono improves ES/EM by 2.40/1.67 points, while two-column layouts cause EM to drop from 11.11 to 10.00.

This divergence stems from varying visual signal priorities. RepoBench-P is constrained by global structure; it relies on spatial cues like indentation rather than character-level resolution, making it font-insensitive but vulnerable to layout-induced noise. In contrast, LCC performance is limited by visual fidelity for local continuation. Monospaced fonts enhance surface-form reconstruction, whereas dense layouts introduce character misrecognition, disproportionately harming the strict Exact Match (EM) metric.

\begin{findingbox}
\textbf{Finding 4}: Visual code compression is governed by a fundamental coverage-fidelity trade-off: it excels in global-dependency tasks by preserving context coverage, but faces a fidelity bottleneck in exactness-critical tasks. While textual code compression pays with structural coverage loss, visual code compression pays with symbolic fidelity loss.
\end{findingbox}

\section{Discussion and Threats to Validity}
\label{Discussion and threats to Validity}

\subsection{Disscussion}

Our experimental results demonstrate the distinct advantages of visual code compression across various long-context scenarios. 
To further understand the underlying mechanisms and practical implications of these findings, this section synthesizes our observations into a theoretical framework, analyzes the trade-offs between different paradigms, and discusses the feasibility of deploying such models in real-world software engineering workflows.
\subsubsection{The Coverage--Fidelity Trade-off: Implications for Choosing Compression Paradigms}
Our empirical study identifies a fundamental mechanism governing long-context code understanding: the \textit{coverage--fidelity trade-off}.
Textual code comression operate by discretely pruning context, often leading to \textit{coverage loss} where omitting prerequisite constraints irreversibly fragments code logic.
In contrast, visual code compression preserves global context via continuous mapping but faces a \textit{fidelity bottleneck} bounded by resolution.
This distinction dictates paradigm suitability based on signal distribution: visual code compression dominates in macro-level reasoning (e.g., summarization) where task-critical signals are sparsely dispersed, whereas textual code comression retains a competitive edge in micro-level tasks requiring high symbolic precision.
Essentially, the choice entails balancing the risk of structural incompleteness against symbolic ambiguity.

\subsubsection{Cost-Efficiency and Interactive Feasibility}
Beyond accuracy, the shift to visual processing introduces a transformative advantage in deployment feasibility.
Achieving high fidelity in selective filtering (e.g., LongCodeZip) typically necessitates expensive model-based metrics, incurring substantial token and latency overheads.
Conversely, LongCodeOCR shifts the computational burden from heavy LLM inference to lightweight graphical rendering.
This paradigm eliminates compression-stage token consumption and drastically reduces preprocessing latency.
Such efficiency unlocks the feasibility of million-token context analysis in resource-constrained, real-time environments such as IDE plugins that remain computationally prohibitive for heavy, model-driven filtering pipelines.

\subsubsection{Future Directions}
\label{sec:future_directions}
Our findings suggest several promising directions for improving the reliability and practicality of visual code compression for long-context code understanding.
First, future work can mitigate symbol-level fidelity bottlenecks by exploring resolution-adaptive rendering and symbol-aware layouts, and by incorporating lightweight verification or correction mechanisms to reduce the impact of OCR noise on exactness-critical tasks.
Second, hybrid pipelines that combine global-preserving visual views with selective filtering-based compression may better balance the coverage--fidelity trade-off under tight context and latency budgets.

\subsection{Threats to Validity}

While our empirical study provides systematic evidence for the efficacy of visual code compression, several factors may impact the validity of our findings. 
Following the standard guidelines for empirical software engineering, we identify and discuss potential threats to validity in terms of internal, external, and construct validity, along with the mitigation strategies we employed.
\subsubsection{Internal Validity.}
A potential threat to internal validity concerns the fairness of comparisons across modalities.
To ensure a rigorous baseline, we match the compression ratios between visual code compression and textual code  compression across all benchmarks.
Moreover, to reduce confounding effects related to model family and pretraining, we compare Qwen3-VL-8B with its matched text-only counterpart, Qwen3-8B.
Because these models are developed within the same model family and likely share broadly similar pretraining data sources, this controlled comparison helps isolate the effect of visual structural representations from differences in model lineage and data exposure.

\subsubsection{External Validity.} 
External validity concerns the generalizability of our findings to other tasks and models.
We mitigate this threat by evaluating our approach on four representative long-context code benchmarks spanning code summarization, code question answering, and code completion.
Moreover, we consider two distinct VLM backbones, Qwen3-VL-8B and Glyph (a specialized 9B VLM), to reduce concerns that the observed benefits of visual compression are specific to a single model.
Finally, our RQ3 results indicate robustness across practical rendering configurations, including font types and spatial layouts.

\subsubsection{Construct Validity.} 
Construct validity concerns whether the benchmark metrics faithfully reflect the intended notion of task correctness, rather than over-emphasizing textual surface forms.
For instance, the EM metric in the LCC task is highly sensitive to symbol-level precision and may disproportionately penalize visual pipelines when OCR introduces minor character-level noise.
To mitigate potential metric-driven bias, we strictly follow each benchmark's standardized evaluation protocol and report the metric(s) it specifies.
In particular, for LCC we report both strict EM and a softer similarity measure such as ES, as prescribed by the benchmark.
This reporting helps avoid over-reliance on a single strict criterion and supports our discussion of the \textit{coverage--fidelity trade-off} between broader context coverage and symbol-level fidelity.

\section{Related Work}\label{sec:related}

\subsection{Long-Context Code Benchmarks and Tasks}
\label{sec:benchmarks}

Code evaluation benchmarks have evolved from standalone function synthesis (typified by
HumanEval~\cite{chen2021evaluating}) to repository-level tasks that require reasoning over
cross-file constraints. This shift has motivated long-context benchmarks that assess evidence
localization and reasoning consistency under extensive inputs.

In \textbf{cross-file completion}, RepoBench~\cite{liu2023repobench},
CrossCodeEval~\cite{ding2023crosscodeeval}, and RepoEval~\cite{zhang2023repocoder} move beyond local
code generation to evaluate project-level context utilization, emphasizing the need to identify
dispersed dependencies while controlling data leakage and memorization.

In \textbf{long-code understanding and maintenance}, RepoQA~\cite{liu2024repoqa} and
LongCodeU~\cite{li2025longcodeu} focus on evidence localization and inter-unit relation
understanding. LongCodeBench~\cite{rando2025longcodebench} further scales contexts to the
million-token level to study degradation under extreme length. In parallel,
SWE-bench~\cite{jimenez2023swe} and Long Code Arena~\cite{bogomolov2024long} ground evaluation in
realistic workflows such as issue resolution and codebase maintenance.

\subsection{Efficient Long-Context Architectures}
\label{subsec:long-context-modeling}

The primary bottleneck in processing massive code repositories lies in the quadratic complexity
of self-attention~\cite{vaswani2017attention}. To reduce the prohibitive compute and memory costs
of extending context windows, prior work largely follows two directions: improving architectural
efficiency and enabling length extrapolation.

\textbf{Architectural Efficiency.}
Early studies explored sparse attention patterns~\cite{beltagy2020longformer,zaheer2020big} and
kernel-based linearization~\cite{wang2020linformer,choromanski2020rethinking}. System-level
innovations such as FlashAttention~\cite{dao2022flashattention} further enable efficient exact
attention via IO-aware tiling. Compressive and recurrent architectures, including
Transformer-XL~\cite{dai2019transformer} and Infini-attention~\cite{munkhdalai2024leave}, maintain
bounded memory states to capture long-range dependencies without scaling computation linearly with
history length.

\textbf{Extrapolation and Utilization.}
Robust positional extrapolation is critical for effectiveness beyond training length.
ALiBi~\cite{press2021train} and RoPE scaling strategies, including YaRN~\cite{peng2309yarn} and
LongRoPE~\cite{ding2024longrope}, extend effective context windows post-training. However, longer
nominal windows alone do not guarantee reliable utilization; recent work emphasizes data curation
and efficient fine-tuning paradigms, including LongLoRA~\cite{chen2023longlora} and
LongAlign~\cite{bai2024longalign}, to improve models' ability to consistently use critical
information in long contexts~\cite{zhao2025vtcbench}.

\subsection{Context Compression Paradigms}
\label{subsec:context-compression-paradigms}

Context compression mitigates the computational burden of long-context modeling by increasing
information density under a constrained context budget. Prior work largely follows two paradigms:
selective filtering-based compression and cross-modal visual compression.

\textbf{Selective filtering-based compression.}
General-purpose compressors such as LLMLingua~\cite{pan2024llmlingua,jiang2023llmlingua} and
Selective Context~\cite{li2023compressing} leverage signals including perplexity to identify and
remove low-utility content. For source code, however, token removal must respect program
structure; otherwise it may violate syntactic well-formedness and disrupt cross-file dependency
chains. Code-oriented methods therefore incorporate structure-aware constraints.
DietCode~\cite{zhang2022diet} and SlimCode~\cite{wang2024natural} utilize syntactic structure and
program dependencies to guide selective filtering while maintaining structural fidelity. At
repository scale, LongCodeZip~\cite{shi2025longcodezip} provides a systematic pipeline for
compressing distributed code contexts and constitutes a strong representative of text-based code
compression.

\textbf{Cross-modal visual compression.}
Vision--text compression (VTC) exploits the visual modality by rendering text into images and
encoding them into a fixed number of visual tokens through vision encoders.
VIST~\cite{xing2025vision} and VisInContext~\cite{wang2024leveraging} show that visual
representations of low-salience context can reduce token costs while maintaining downstream
performance. Related ``optical computing'' lines of work, including
DeepSeek-OCR~\cite{wei2025deepseek}, Glyph~\cite{cheng2025glyph}, and
AgentOCR~\cite{feng2026agentocr}, further demonstrate the feasibility of extreme compression for
OCR and long-horizon agent histories by leveraging two-dimensional layouts and vision--language models.

\section{Conclusion}
\label{Conclusion}
In this paper, we conducted a systematic empirical study to evaluate the efficacy and applicability boundaries of visual code compression relative to textual code compression for long-context code understanding. Under a controlled and identical experimental protocol, we benchmarked LongCodeOCR against representative baselines, including LongCodeZip, across a diverse suite of long-context software engineering tasks.
Our results characterize a fundamental coverage--fidelity trade-off inherent in these paradigms: visual code compression favors retaining higher context coverage to support long-range constraints and global evidence integration, thereby exhibiting more stable behavior in scenarios requiring broad semantic coverage. Conversely, textual code compression maintains textual symbolic precision and remains competitive in settings that are highly sensitive to local surface forms under high compression strengths. Furthermore, our analysis indicates that the primary risk of textual code compression stems from coverage loss, whereas the bottleneck of visual code compression is manifested as fidelity degradation affecting fine-grained symbol recognition.
Finally, we discuss the implications and offer guidelines on when to prioritize visual compression and how to navigate the trade-offs between compression strength and rendering configurations.


\bibliographystyle{ACM-Reference-Format}
\bibliography{ref}

\appendix

\end{document}